%% file: Artigo.tex
\documentclass[conference]{IEEEtran}
\IEEEoverridecommandlockouts

\usepackage{cite}
\usepackage{amsmath,amssymb,amsfonts}
\usepackage{algorithmic}
\usepackage{graphicx}
\usepackage{textcomp}
\usepackage{xcolor}

\usepackage[colorlinks=true, allcolors=blue]{hyperref}

\usepackage{todonotes}
\usepackage{multirow}
\newcounter{xaviercounter}

\newcommand{\ourlab}{the Center Inria of Bordeaux University }

\usepackage{listings}

\usepackage{csquotes}

\lstdefinestyle{pythonstyle}{
    language=Python,
    basicstyle=\ttfamily\small,
    keywordstyle=\color{blue}\bfseries,
    stringstyle=\color{green!70!black},
    commentstyle=\color{gray},
    numberstyle=\tiny\color{gray},
    numbersep=5pt,
    showstringspaces=false,
    breaklines=true,
    frame=single,
    tabsize=4
}
\usepackage{enumitem}

\def\BibTeX{{\rm B\kern-.05em{\sc i\kern-.025em b}\kern-.08em
    T\kern-.1667em\lower.7ex\hbox{E}\kern-.125emX}}

\begin{document}

\title{Deploying Open-Source Large Language Models: \\A performance Analysis
\thanks{Preprint version}
}


\author{\IEEEauthorblockN{Yannis Bendi-Ouis}
\IEEEauthorblockA{\textit{Mnemosyne} \\
\textit{Inria}\\
Bordeaux, France \\
yannis.bendi-ouis@inria.fr}
\and
\IEEEauthorblockN{Dan Dutartre}
\IEEEauthorblockA{\textit{SED} \\
\textit{Inria}\\
Bordeaux, France \\
dan.dutartre@inria.fr}
\and
\IEEEauthorblockN{Hinaut Xavier}
\IEEEauthorblockA{\textit{Mnemosyne} \\
\textit{Inria}\\
Bordeaux, France \\
xavier.hinaut@inria.fr}
}
\maketitle

\begin{abstract}
Since the release of ChatGPT in November 2022, large language models (LLMs) have seen considerable success, including in the open-source community, with many open-weight models available. However, the requirements to deploy such a service are often unknown and difficult to evaluate in advance.
To facilitate this process, we conducted numerous tests at \ourlab. In this article, we propose a comparison of the performance of several models of different sizes (mainly Mistral and LLaMa) depending on the available GPUs, using vLLM, a Python library designed to optimize the inference of these models.
Our results provide valuable information for private and public groups wishing to deploy LLMs, allowing them to evaluate the performance of different models based on their available hardware. This study thus contributes to facilitating the adoption and use of these large language models in various application domains.
\end{abstract}

\section{Introduction}
Following the release of ChatGPT by OpenAI in November 2022 \cite{achiam2023gpt}, large language models (LLMs) \cite{zhao2023survey} have sparked great interest in the private sector, driving numerous companies to invest in LLM-based services, setting off a wave of innovation and competition. We can notably mention tech giants like Microsoft with Copilot \cite{peng2023impact}, Google with Gemini \cite{team2024gemini}, Meta with Llama \cite{touvron2023llama} and Anthropic with Claude. This explosion of interest has transformed the landscape of artificial intelligence, pushing companies and researchers to explore the vast potential of LLMs. However, training and inference of such models remain inaccessible to the general public, requiring considerable computational power and high-quality data. For instance, Meta AI's acquisition of 350,000 NVIDIA H100 GPUs in January 2024, estimated at \$9 billion, underscores the immense financial and technological investment required for cutting-edge research like their LLaMa-3 model, which was trained on an astronomical 14 × 10¹² tokens. 

Recognizing the potential monopolization of this technology, leading corporations are actively lobbying governments to regulate LLMs, citing the risks and potential dangers of their malicious use. Their proposed measures ranging from restricting model training beyond certain computational thresholds to government control of GPU resources, with the possibility of remote deactivation \cite{openAIstructure}. Despite these efforts to consolidate power, it is imperative that such technologies are not confined solely to a few dominant actors. One critical aspect that cannot be overlooked is the ethical implications of data selection and model training. The individuals or organizations responsible for training LLMs have significant control over the datasets used, which inherently contain biases reflecting their perspectives, values, and priorities. This power enables them to shape the narrative, rewrite history, or propagate their own agendas — potentially leading to the distortion of information and the dissemination of propaganda. In an era of "post-truths" and "alternative facts", the concentration of this power in the hands of a few entities could lead to the manipulation of public opinion at a large scale, further entrenching existing biases and reinforcing misinformation. Transparency emerges as a critical solution, enabling external entities to scrutinize the integrity and security of proposed models, ensuring that they remain truthful and aligned with the broader interests of society rather than serving narrow agendas. While many companies are reluctant to embrace open-source initiatives, some pioneers like Meta (with their LLaMA series) \cite{touvron2023llama}, Mistral AI \cite{jiang2024mixtral}, and DeepSeek \cite{liu2024deepseek} \cite{guo2025deepseek} have made significant strides in promoting open-weight models. These efforts allow for greater democratization of AI technology, empowering diverse groups — both public and private — to deploy powerful models independently. 

The availability of open-source models has opened new avenues for innovation, enabling researchers and developers to experiment with LLMs without relying on proprietary systems. This shift is particularly significant in the context of smaller, more efficient models that have gained traction as viable alternatives to larger counterparts. These smaller models, often referred to as "light" or "distilled" models, require fewer computational resources and are easier to deploy, making them accessible to a broader audience. Nevertheless, even for smaller models (ranging from 7B to 30B parameters), the computational power required was still significant. As a result, a portion of the open-source community turned its attention to quantizing model weights and creating open-source packages or services designed to run these models locally. However, even with these advancements, deploying LLMs at scale remains challenging. While serving a model to a single user is straightforward, scaling up to tens, hundreds, or thousands of simultaneous users demands meticulous planning and optimization. This complexity highlights the importance of thorough testing and evaluation, as oversights in resource allocation or infrastructure design can lead to inefficiencies and reduced performance. In this context, we conducted several tests at \ourlab concerning the deployment of such models.

\section{Objectives}
The main objective of our study is to address the security and confidentiality concerns raised by the increasing use of proprietary LLM solutions - such as ChatGPT - by students and researchers at \ourlab. Indeed, a huge part of our students use these tools to help them in their daily work, whether for writing, programming, proofreading articles, or brainstorming. For instance, leveraging Retrieval-Augmented Generation (RAG) systems \cite{gao2023retrieval} \cite{edgetrinh2024graphrag} has become a valuable tool in both academic and practical settings. RAG combines language models with external knowledge sources, enabling users to engage in detailed explorations of specific datasets or domains, such as historical documents or scientific papers. This application is particularly beneficial for students and researchers, as it enhances their ability to analyze and understand complex information.

However, the use of these proprietary solutions raises serious security and confidentiality issues. They do not guarantee the confidentiality of data, and the private interests behind them can potentially use them for commercial purposes, training, or even industrial espionage. This last point is particularly concerning for a research center like \ourlab, which must ensure the confidentiality of its employees' research work and is in direct competition with the companies offering these proprietary solutions. It is therefore crucial for our lab to propose alternative solutions and preserve its digital sovereignty. 

\section{Prerequisites}
\subsection{Skills}
To deploy an LLM on a GPU, certain knowledge and skills are required in Linux and Python development, as well as a strong curiosity about existing models and quantification. Although understanding the internal workings of Transformers is not necessary, it can be an asset. The required skills include the ability to update CUDA drivers (version 12 minimum), install a version of Python (minimum 3.9), install Python dependencies, make HTTP requests, and choose the right model for your use case, quantified or not depending on your resources.

\subsection{Hardware}
We conducted our tests on the Plafrim computing server, equipped with two types of GPUs:

\begin{itemize}
    \item NVIDIA V100 16 GB
    \item NVIDIA A100 40 GB
\end{itemize}

\subsection{Software}
We used vLLM \cite{kwon2023efficient}, a Python library designed to optimize the inference of these models. This library requires at least the prior installation of Python 3.9 and CUDA 12. The advantage of vLLM over other solutions is its ability to handle multiple requests simultaneously, without a queue and without a linear increase in computation time depending on the number of requests, but rather logarithmic. However, depending on the available hardware, other solutions can be considered. Notably, tensorRT-LLM offers excellent performance with NVIDIA GPUs, and lllama.cpp provides remarkable performance on Macs equipped with M1, M2, or M3 chips.

\subsection{Quantification}
Some models can be very large, making it particularly difficult to load them into the available hardware - limited by its VRAM. To address this problem, one of the best solutions is to quantify our models. Instead of writing the values of our weights on 16 or 32 bits, we can accept a slight loss of precision and write them on 4 or 8 bits. This loss has been evaluated several times, and although it varies depending on the models and quantification methods used, we can affirm that it is negligible up to 6-bits and acceptable up to 4-bits. However, for a number of parameters greater than 70 billion, the models are robust enough to allow quantification below 4-bits while maintaining good coherence. Among the different quantification methods \cite{rajput2024benchmarking}, we can mention AWQ \cite{lin2024awq}, GPTQ \cite{frantar2022gptq}, and GGUF (llama.cpp).

\section{Experimentation}
In this study, we seek to determine the maximum load of simultaneous requests that a server equipped with V100 16 GB or A100 40 GB GPUs can support, depending on the LLM used. For this, we conducted tests by progressively increasing the number of simultaneous requests and the size of the prompts, until reaching the maximum load. For each request, we measured the time required to generate 100 tokens. And, for each model and GPU, we measured the memory load, execution speed, and number of tokens per second depending on the number of simultaneous requests and the maximum context size.

We chose to focus mainly on the models proposed by Mistral AI \cite{jiang2023mistral} \cite{jiang2024mixtral}, due to their diversity, popularity, and skills. We also appreciate their performance on European languages, particularly French. Additionally, their Mixture of Experts \cite{jiang2024mixtral} architecture allows for computational savings during inference, by selecting only a part of the model's weights at each step, which also reduces energy consumption. Moreover, we included the LLaMa-3-70B \cite{touvron2023llama} model from Meta, which achieves performance comparable to GPT-4 with its 70 billion parameters. This model size seems to be a good compromise between size and performance, justifying its inclusion in our study. Thus, we tested the following models: Mistral-7B, Codestral-22b, Mixtral-8x7b, Mixtral-8x22b and LLaMa-3-70B.

\newpage

\section{Results}

\subsection{Mistral 7B on 2 V100 16 GB}
\begin{table}[ht!]
\centering
\begin{tabular}{|c|c|c|c|c|c|c|c|}
\hline
\multirow{2}{*}{Requests} & \multicolumn{7}{c|}{Number of tokens} \\
\cline{2-8}
 & 31 & 63 & 119 & 296 & 480 & 822 & 2193 \\
\hline
1 & 1.8 & 1.8 & 1.9 & 1.9 & 1.9 & 2.1 & 2.3 \\
\hline
2 & 2.1 & 2.1 & 2.0 & 2.2 & 2.3 & 2.6 & 2.8 \\
\hline
4 & 2.2 & 2.3 & 2.1 & 2.6 & 2.5 & 2.8 & 3.7 \\
\hline
8 & 2.4 & 2.4 & 2.5 & 2.7 & 3.0 & 3.5 & 5.9 \\
\hline
16 & 2.9 & 2.9 & 3.0 & 3.8 & 4.2 & 5.2 & 9.2 \\
\hline
32 & 4.2 & 4.2 & 4.5 & 5.4 & 6.9 & 8.8 & 19.0 \\
\hline
64 & 6.7 & 7.1 & 7.7 & 9.8 & 11.9 & 17.1 & 36.0 \\
\hline
128 & 10.6 & 10.4 & 11.5 & 16.2 & 24.4 & 33.3 & 72.1 \\
\hline
\end{tabular}
\hspace{1cm}
\caption{Time (s) to resolve the request with Mistral 7B on 2 V100 16 GB}
\end{table}

\subsection{Codestral 22B AWQ 4-bits on 1 A100 40 GB}
\begin{table}[ht!]
\centering
\begin{tabular}{|c|c|c|c|c|c|c|c|}
\hline
\multirow{2}{*}{Requests} & \multicolumn{7}{c|}{Number of tokens} \\
\cline{2-8}
 & 31 & 63 & 119 & 296 & 480 & 822 & 2193 \\
\hline
1 & 2.3 & 2.3 & 2.4 & 2.5 & 2.6 & 2.6 & 3.0 \\
\hline
2 & 2.3 & 2.4 & 2.5 & 2.7 & 2.7 & 2.8 & 3.5 \\
\hline
4 & 2.4 & 2.5 & 2.6 & 2.8 & 3.0 & 3.4 & 4.8 \\
\hline
8 & 2.6 & 2.7 & 2.8 & 3.2 & 3.7 & 4.5 & 7.4 \\
\hline
16 & 3.0 & 3.2 & 3.4 & 4.2 & 5.0 & 6.4 & 12.3 \\
\hline
32 & 4.5 & 4.8 & 5.4 & 6.7 & 8.4 & 11.4 & 23.1 \\
\hline
64 & 7.9 & 8.5 & 9.3 & 12.3 & 15.8 & 21.6 & 47.7 \\
\hline
128 & 14.3 & 15.4 & 17.6 & 24.2 & 29.9 & 46.4 & 96.2 \\
\hline
\end{tabular}
\hspace{1cm}
\caption{Time (s) to resolve the request with Codestral 22B AWQ 4-bits on 1 A100 40 GB}
\end{table}

\subsection{Codestral 22B GPTQ 8-bits on 2 V100 16 GB}
\begin{table}[ht!]
\centering
\begin{tabular}{|c|c|c|c|c|c|c|c|}
\hline
\multirow{2}{*}{Requests} & \multicolumn{7}{c|}{Number of tokens} \\
\cline{2-8}
 & 31 & 63 & 119 & 296 & 480 & 822 & 2193 \\
\hline
1 & 3.1 & 3.2 & 3.2 & 3.3 & 3.4 & 3.7 & 4.3 \\
\hline
2 & 3.3 & 3.4 & 3.4 & 3.6 & 4.0 & 4.4 & 5.8 \\
\hline
4 & 3.7 & 3.8 & 3.9 & 4.4 & 4.8 & 5.6 & 8.8 \\
\hline
8 & 4.8 & 5.1 & 5.3 & 6.0 & 6.8 & 8.8 & 15 \\
\hline
16 & 7.1 & 7.5 & 7.9 & 9.8 & 11.7 & 14.3 & 27.5 \\
\hline
32 & 10.4 & 10.9 & 11.9 & 15.3 & 19.0 & 24.6 & 53.8 \\
\hline
64 & 15.5 & 17.0 & 18.7 & 25.9 & 32.1 & 43.9 & 108.2 \\
\hline
128 & 21.7 & - & - & - & - & - & - \\
\hline
\end{tabular}
\hspace{1cm}
\caption{Time (s) to resolve the request with Codestral 22B GPTQ 8-bits on 2 V100 16 GB}
\end{table}

The first thing we can notice is that the larger the context size, the slower the model is at generating 100 tokens. This is expected and is related to its complexity, which grows quadratically with the context size. Moreover, as we can see with the Mixtral \cite{jiang2024mixtral} and Llama \cite{touvron2023llama} models, just because a model can be loaded does not mean it can be used for any context size. The context size has a quadratic cost in RAM (or VRAM), which adds to the model size, potentially significantly increasing memory requirements. On the other hand, we can also observe that we do not have a linear loss of efficiency with the number of simultaneous requests: the time required to respond to a request does not double when the number of simultaneous requests doubles.

\subsection{Codestral 22B on 2 A100 40 GB}
\begin{table}[ht!]
\centering
\begin{tabular}{|c|c|c|c|c|c|c|c|}
\hline
\multirow{2}{*}{Requests} & \multicolumn{7}{c|}{Number of tokens} \\
\cline{2-8}
 & 31 & 63 & 119 & 296 & 480 & 822 & 2193 \\
\hline
1 & 2.3 & 2.3 & 2.4 & 2.4 & 2.5 & 2.6 & 2.8 \\
\hline
2 & 2.3 & 2.3 & 2.4 & 2.5 & 2.6 & 2.7 & 3.3 \\
\hline
4 & 2.4 & 2.4 & 2.5 & 2.7 & 2.8 & 3.1 & 4.2 \\
\hline
8 & 2.5 & 2.6 & 2.8 & 3.1 & 3.4 & 4.1 & 6.3 \\
\hline
16 & 2.8 & 2.9 & 3.2 & 3.8 & 4.4 & 5.6 & 10.2 \\
\hline
32 & 3.3 & 3.7 & 4.0 & 5.2 & 6.4 & 8.8 & 18.1 \\
\hline
64 & 4.3 & 4.6 & 5.7 & 8.0 & 10.5 & 15.5 & 36.8 \\
\hline
128 & 6.8 & 7.8 & 9.4 & 14.5 & 19.6 & 32.9 & 71.1 \\
\hline
\end{tabular}
\hspace{2cm}
\caption{Time (s) to resolve the request with Codestral 22B on 2 A100 40 GB}
\end{table}

\subsection{Mixtral 8x7B AWQ 4-bits on 2 A100 40 GB}
\begin{table}[ht!]
\centering
\begin{tabular}{|c|c|c|c|c|c|c|c|}
\hline
\multirow{2}{*}{Requests} & \multicolumn{7}{c|}{Number of tokens} \\
\cline{2-8}
 & 31 & 63 & 119 & 296 & 480 & 822 & 2193 \\
\hline
1 & 3.1 & 3.2 & 3.6 & 3.4 & 3.5 & 3.5 & 4.1 \\
\hline
2 & 3.3 & 3.3 & 3.5 & 3.5 & 3.8 & 3.8 & 4.7 \\
\hline
4 & 3.5 & 3.6 & 3.5 & 4.3 & 4.1 & 4.6 & 6.2 \\
\hline
8 & 3.8 & 3.8 & 4.0 & 4.3 & 4.9 & 5.7 & 9.3 \\
\hline
16 & 4.3 & 4.6 & 4.9 & 6.0 & 6.7 & 8.5 & 15.5 \\
\hline
32 & 6.0 & 6.4 & 7.6 & 8.7 & 10.9 & 14.2 & - \\
\hline
64 & 10.0 & 10.5 & 11.6 & 15.5 & - & - & - \\
\hline
128 & 18.5 & 19.6 & 21.5 & - & - & - & - \\
\hline
\end{tabular}
\hspace{1cm}
\caption{Time (s) to resolve the request with Mixtral 8x7B AWQ 4-bits on 2 A100 40 GB}
\end{table}

\subsection{Mixtral 8x22B AWQ 4-bits on 4 A100 40 GB}
\begin{table}[ht!]
\centering
\begin{tabular}{|c|c|c|c|c|c|c|c|}
\hline
\multirow{2}{*}{Requests} & \multicolumn{7}{c|}{Number of tokens} \\
\cline{2-8}
 & 31 & 63 & 119 & 296 & 480 & 822 & 2193 \\
\hline
1 & 6.0 & 6.0 & 6.3 & 6.8 & 7.0 & 7.6 & 10.9 \\
\hline
2 & 7.2 & 7.1 & 7.4 & 8.4 & 8.7 & 10.3 & 16.7 \\
\hline
4 & 8.0 & 8.1 & 8.7 & 10.3 & 11.9 & 14.2 & 21.3 \\
\hline
8 & 9.0 & 9.4 & 10.0 & 13.0 & 16.7 & 19.4 & 36.2 \\
\hline
16 & 11.0 & 12.2 & 13.2 & 21.1 & 26.5 & 31.9 & 66.4 \\
\hline
32 & 16.0 & 17.6 & 22.6 & 33.6 & 37.7 & 56.7 & - \\
\hline
64 & 28.0 & 31.8 & 35.9 & 56.0 & 71.7 & - & - \\
\hline
128 & 52.0 & 55.3 & 67.0 & 111.7 & - & - & - \\
\hline
\end{tabular}
\hspace{1cm}
\caption{Time (s) to resolve the request with Mixtral 8x22B AWQ 4-bits on 4 A100 40 GB}
\end{table}

\vspace{-0.3cm}

\subsection{LLaMa-3 70B AWQ 4-bits on 2 A100 40 GB}
\begin{table}[ht!]
\centering
\begin{tabular}{|c|c|c|c|c|c|c|c|}
\hline
\multirow{2}{*}{Requests} & \multicolumn{7}{c|}{Number of tokens} \\
\cline{2-8}
 & 21 & 51 & 97 & 240 & 398 & 703 & 1848 \\
\hline
1 & 3.6 & 3.7 & 3.7 & 3.9 & 4.0 & 4.2 & 4.8 \\
\hline
2 & 3.7 & 3.7 & 3.9 & 4.1 & 4.2 & 4.5 & 5.8 \\
\hline
4 & 3.8 & 4.0 & 4.1 & 4.4 & 4.7 & 5.4 & 7.9 \\
\hline
8 & 4.3 & 4.5 & 4.8 & 5.1 & 5.9 & 7.4 & 12.5 \\
\hline
16 & 4.9 & 5.2 & 5.7 & 6.9 & 8.3 & 10.8 & 21.4 \\
\hline
32 & 7.6 & 8.2 & 8.7 & 11.1 & 13.7 & 19.6 & 40.9 \\
\hline
64 & 12.9 & 13.9 & 15.2 & 20.3 & 25.8 & 37.5 & - \\
\hline
128 & 23.2 & - & - & - & - & - & - \\
\hline
\end{tabular}
\hspace{1cm}
\caption{Time (s) to resolve the request with LLaMa-3 70B AWQ 4-bits on 2 A100 40 GB}
\end{table}

\begin{figure*}
    \hspace{-20pt}
    \centering
    \includegraphics[width=1\linewidth]{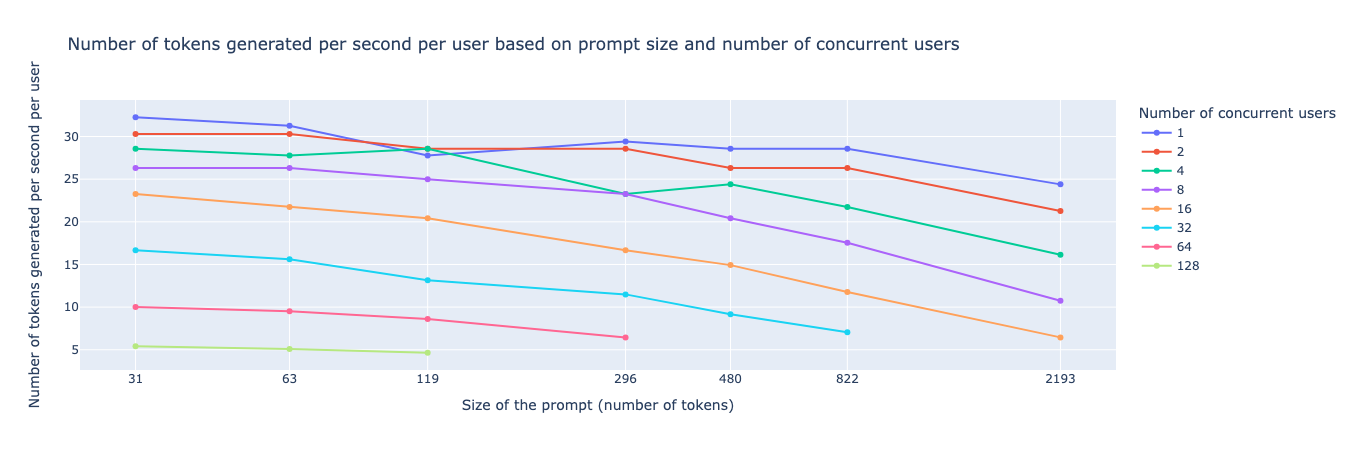}
    \caption{Number of tokens generated per second and per user for Mixtral 8x7B AWQ 4-bits on 2 A100 40GB}
\end{figure*}

\begin{figure*}
    \hspace{-20pt}
    \centering
    \includegraphics[width=1\linewidth]{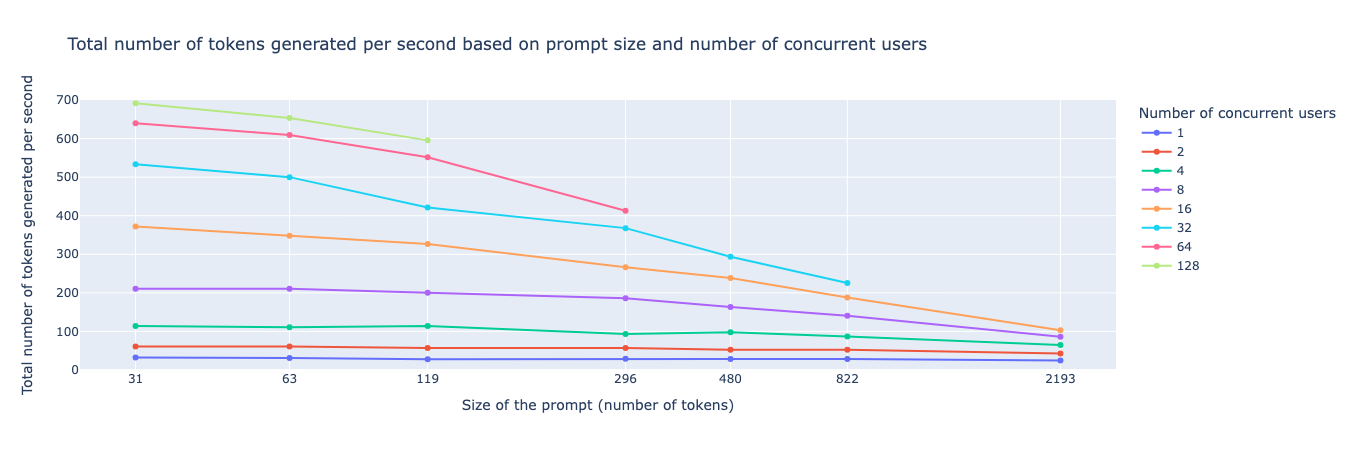}
    \caption{Total number of token generated per second for Mixtral 8x7B AWQ 4-bits on 2 A100 40GB}
\end{figure*}

However, this seems less true when the request size exceeds a certain threshold.

Finally, we can notice that although the cost of these GPUs is far from trivial (V100 16GB $\approx$ \$5000, A100 40GB  $ \approx $ \$8500), it is not necessary to possess an exorbitant quantity to successfully run a local alternative to ChatGPT or other proprietary solutions. In fact, with two A100 40GB GPUs (or a single 80GB GPU), it is already possible to run LLaMa-3-70B or Mixtral 8x7B under very good conditions. According to numerous benchmarks and user reviews, these models are serious contenders to GPT-4. 

To a lesser extent, it is possible to host smaller models (in the range of 7 to 30B) and achieve extremely impressive generation speeds, especially if requests are parallelized. For example, we can observe in Figure 1 the number of tokens generated per second and per user as a function of the number of simultaneous requests and the size of the prompt, and in Figure 2 the total number of tokens generated per second (for all users combined) as a function of the number of simultaneous requests and the size of the prompts. Thus, we can observe that for a model like Mixtral 8x7B — which has 49B parameters and only 12-13B active parameters at any given time (MoE architecture) \cite{jiang2024mixtral} — we can achieve up to 700 tokens/second for 128 simultaneous requests with a nearly zero context (30 tokens), and we can obtain an inference speed of just under 20 tokens per second for 20 simultaneous users.

\section{Discussion \& Perspective}

In this article, we have presented a comparative study of the performance of several large language models (LLMs) based on available hardware resources. Our experimental results demonstrate that deploying open-source LLMs such as Mistral \cite{jiang2023mistral} \cite{jiang2024mixtral} and LLaMA \cite{touvron2023llama} on V100 and A100 GPUs is feasible and highly efficient. The performance of these models, particularly when equipped with advanced architectures like the Mixture-of-Experts (MoE), rivals or even surpasses proprietary solutions like ChatGPT \cite{achiam2023gpt}. One key observation from our experiments is the scalability of these models when serving multiple simultaneous users. While the time required to respond to a request does not linearly increase with the number of requests, there is a threshold beyond which performance begins to degrade. This highlights the need for careful resource allocation and infrastructure design when deploying LLMs at scale. Looking ahead, the integration of newer GPUs like NVIDIA's H100 could further enhance the performance of these models. With more powerful hardware and larger GPU clusters, organizations could potentially deploy even larger models, such as LLaMA-3-405B or DeepSeek-V3, with greater efficiency and scalability. Additionally, advancements in quantization techniques and model compression could make it possible to run these models on less powerful hardware, broadening their accessibility.

They also highlight the importance of transparency and digital sovereignty, enabling public and private groups to deploy open-source models without relying on proprietary solutions. One of the most compelling reasons for organizations to deploy their own LLMs locally is the snowball effect it creates. When data is shared with external providers, it contributes to the improvement of their models, giving them a competitive edge. However, by deploying and refining our own models, we can harness this data to improve our own systems, ensuring that the benefits of technological advancements remain within our organization. This not only enhances our capabilities but also strengthens our position in the competitive landscape.

The recent emergence of state-of-the-art (SOTA) open-weight models like DeepSeek V3 and R1 by the end of 2024 further underscores the viability of local deployment. These models, particularly with their Mixture of Experts (MoE) architecture, demonstrate that high performance can be achieved without exorbitant computational costs, provided sufficient VRAM is available. This development is a game-changer, as it shows that open-source models can now rival or even surpass proprietary solutions in terms of performance and efficiency. For organizations that may not have access to the latest hardware, "distilled" versions of these models, such as the "R1 distilled" versions based on Llama and Qwen2.5, offer a practical alternative. These "distilled" models retain much of the performance of their larger counterparts while being more resource-efficient, making them accessible to a wider range of users. We strongly encourage research centers, universities, and businesses to take the initiative in deploying their own LLM solutions. By leveraging open-source (or open-weight) models, organizations can reduce their dependence on external providers and gain greater control over their data and technological infrastructure. This not only enhances data security and confidentiality but also fosters innovation and self-reliance. Moreover, local deployment of LLMs can be tailored to the specific needs of the organization, whether it be for research, education, or business applications. For instance, universities can provide students and researchers with powerful tools for academic work, while businesses can develop customized solutions for their unique challenges. Looking ahead, the continuous improvement of open-source models and the increasing availability of high-performance hardware will further democratize access to LLMs. We anticipate that more organizations will adopt local deployment strategies, leading to a more diverse and competitive ecosystem. This, in turn, will drive innovation and ensure that the benefits of AI are widely distributed.

\section{Conclusion}

Our study successfully demonstrates the feasibility of deploying open-source LLMs such as Mistral and LLaMA on V100 and A100 GPUs. The results highlight the efficiency and scalability of these models, even under significant computational constraints. During various events, we were able to deploy these models in production and serve a large number of simultaneous users without compromising performance. This real-world validation underscores the effectiveness of our approach and reinforces the potential of open-source AI.

By leveraging local deployment strategies and taking advantage of advancements in hardware and model architecture, organizations can achieve high-performance LLMs while maintaining control over their data and infrastructure. As the field of AI continues to evolve, it is crucial that we prioritize ethical considerations, transparency, and sustainability to ensure that these technologies benefit society as a whole.

\section{Acknowledgment}

We extend our gratitude to \ourlab and the Plafrim Cluster for their support and resources, which made this study possible.

\input{refs.bbl}

\end{document}

%% file: refs.bbl

%% file: Artigo.bbl
\begin{thebibliography}{10}
\providecommand{\url}[1]{#1}
\csname url@samestyle\endcsname
\providecommand{\newblock}{\relax}
\providecommand{\bibinfo}[2]{#2}
\providecommand{\BIBentrySTDinterwordspacing}{\spaceskip=0pt\relax}
\providecommand{\BIBentryALTinterwordstretchfactor}{4}
\providecommand{\BIBentryALTinterwordspacing}{\spaceskip=\fontdimen2\font plus
\BIBentryALTinterwordstretchfactor\fontdimen3\font minus \fontdimen4\font\relax}
\providecommand{\BIBforeignlanguage}[2]{{%
\expandafter\ifx\csname l@#1\endcsname\relax
\typeout{** WARNING: IEEEtran.bst: No hyphenation pattern has been}%
\typeout{** loaded for the language `#1'. Using the pattern for}%
\typeout{** the default language instead.}%
\else
\language=\csname l@#1\endcsname
\fi
#2}}
\providecommand{\BIBdecl}{\relax}
\BIBdecl

\bibitem{achiam2023gpt}
J.~Achiam, S.~Adler, S.~Agarwal, L.~Ahmad, I.~Akkaya, F.~L. Aleman, D.~Almeida, J.~Altenschmidt, S.~Altman, S.~Anadkat \emph{et~al.}, ``Gpt-4 technical report,'' \emph{arXiv preprint arXiv:2303.08774}, 2023.

\bibitem{zhao2023survey}
W.~X. Zhao, K.~Zhou, J.~Li, T.~Tang, X.~Wang, Y.~Hou, Y.~Min, B.~Zhang, J.~Zhang, Z.~Dong \emph{et~al.}, ``A survey of large language models,'' \emph{arXiv preprint arXiv:2303.18223}, 2023.

\bibitem{peng2023impact}
S.~Peng, E.~Kalliamvakou, P.~Cihon, and M.~Demirer, ``The impact of ai on developer productivity: Evidence from github copilot,'' \emph{arXiv preprint arXiv:2302.06590}, 2023.

\bibitem{team2024gemini}
G.~Team, P.~Georgiev, V.~I. Lei, R.~Burnell, L.~Bai, A.~Gulati, G.~Tanzer, D.~Vincent, Z.~Pan, S.~Wang \emph{et~al.}, ``Gemini 1.5: Unlocking multimodal understanding across millions of tokens of context,'' \emph{arXiv preprint arXiv:2403.05530}, 2024.

\bibitem{touvron2023llama}
H.~Touvron, T.~Lavril, G.~Izacard, X.~Martinet, M.-A. Lachaux, T.~Lacroix, B.~Rozi{\`e}re, N.~Goyal, E.~Hambro, F.~Azhar \emph{et~al.}, ``Llama: Open and efficient foundation language models,'' \emph{arXiv preprint arXiv:2302.13971}, 2023.

\bibitem{openAIstructure}
OpenAI, ``Reimagining secure infrastructure for advanced ai,'' \emph{Section: blog}, 2024.

\bibitem{jiang2024mixtral}
A.~Q. Jiang, A.~Sablayrolles, A.~Roux, A.~Mensch, B.~Savary, C.~Bamford, D.~S. Chaplot, D.~d.~l. Casas, E.~B. Hanna, F.~Bressand \emph{et~al.}, ``Mixtral of experts,'' \emph{arXiv preprint arXiv:2401.04088}, 2024.

\bibitem{liu2024deepseek}
A.~Liu, B.~Feng, B.~Xue, B.~Wang, B.~Wu, C.~Lu, C.~Zhao, C.~Deng, C.~Zhang, C.~Ruan \emph{et~al.}, ``Deepseek-v3 technical report,'' \emph{arXiv preprint arXiv:2412.19437}, 2024.

\bibitem{guo2025deepseek}
D.~Guo, D.~Yang, H.~Zhang, J.~Song, R.~Zhang, R.~Xu, Q.~Zhu, S.~Ma, P.~Wang, X.~Bi \emph{et~al.}, ``Deepseek-r1: Incentivizing reasoning capability in llms via reinforcement learning,'' \emph{arXiv preprint arXiv:2501.12948}, 2025.

\bibitem{gao2023retrieval}
Y.~Gao, Y.~Xiong, X.~Gao, K.~Jia, J.~Pan, Y.~Bi, Y.~Dai, J.~Sun, and H.~Wang, ``Retrieval-augmented generation for large language models: A survey,'' \emph{arXiv preprint arXiv:2312.10997}, 2023.

\bibitem{edgetrinh2024graphrag}
\BIBentryALTinterwordspacing
D.~Edge, H.~Trinh, N.~Cheng, J.~Bradley, A.~Chao, A.~Mody, S.~Truitt, and J.~Larson, ``From local to global: A graph rag approach to query-focused summarization,'' 2024. [Online]. Available: \url{https://arxiv.org/abs/2404.16130}
\BIBentrySTDinterwordspacing

\bibitem{kwon2023efficient}
W.~Kwon, Z.~Li, S.~Zhuang, Y.~Sheng, L.~Zheng, C.~H. Yu, J.~Gonzalez, H.~Zhang, and I.~Stoica, ``Efficient memory management for large language model serving with pagedattention,'' in \emph{Proceedings of the 29th Symposium on Operating Systems Principles}, 2023, pp. 611--626.

\bibitem{rajput2024benchmarking}
S.~Rajput and T.~Sharma, ``Benchmarking emerging deep learning quantization methods for energy efficiency,'' in \emph{2024 IEEE 21st International Conference on Software Architecture Companion (ICSA-C)}.\hskip 1em plus 0.5em minus 0.4em\relax IEEE, 2024, pp. 238--242.

\bibitem{lin2024awq}
J.~Lin, J.~Tang, H.~Tang, S.~Yang, W.-M. Chen, W.-C. Wang, G.~Xiao, X.~Dang, C.~Gan, and S.~Han, ``Awq: Activation-aware weight quantization for on-device llm compression and acceleration,'' \emph{Proceedings of Machine Learning and Systems}, vol.~6, pp. 87--100, 2024.

\bibitem{frantar2022gptq}
E.~Frantar, S.~Ashkboos, T.~Hoefler, and D.~Alistarh, ``Gptq: Accurate post-training quantization for generative pre-trained transformers,'' \emph{arXiv preprint arXiv:2210.17323}, 2022.

\bibitem{jiang2023mistral}
A.~Q. Jiang, A.~Sablayrolles, A.~Mensch, C.~Bamford, D.~S. Chaplot, D.~d.~l. Casas, F.~Bressand, G.~Lengyel, G.~Lample, L.~Saulnier \emph{et~al.}, ``Mistral 7b,'' \emph{arXiv preprint arXiv:2310.06825}, 2023.

\end{thebibliography}
